\documentclass{aa} 
\usepackage{graphics}
\begin{document}

\title{Water maser emission in IC~342}


\author{A. Tarchi\inst{1}
        \and
        C. Henkel\inst{1}
        \and
        A. B. Peck\inst{1}
        \and
        K. M. Menten\inst{1}
        }

\offprints{A. Tarchi,
\email{atarchi@mpifr-bonn.mpg.de}}

\institute{Max-Planck-Institut f\"ur Radioastronomie, Auf dem H\"ugel
              69, D-53121 Bonn, Germany}

\date{Received date / Accepted date}

\abstract{The detection of 22\,GHz water vapor emission from IC~342 is 
reported, raising the detection rate among northern galaxies with IRAS
point source fluxes $S_{\rm 100\mu m}$ $>$ 50\,Jy to 16\%. The maser, 
associated with a star forming region $\sim$10--15\arcsec\ west of the nucleus, 
consists of a single 0.5\,km\,s$^{-1}$ wide feature and reaches an isotropic 
luminosity of 10$^{-2}$\,L$_{\odot}$ ($D$=1.8\,Mpc). If the time variability is 
intrinsic, the maser size is $\la1.5\times10^{16}$\,cm ($\la$0.5\,mas) which 
corresponds to a brightness temperature of $\ga$10$^{9}$\,K. The linewidth, 
luminosity, and rapid variability are reminiscent of the 8\,km\,s$^{-1}$ 
super maser in Orion-KL. A velocity shift of 1\,km\,s$^{-1}$ within two weeks
and subsequent rapid fading is explained in terms of a chance alignment of
two dense molecular clouds. Observations at 22 GHz toward Maffei~2 are also reported, yielding a 5$\sigma$ 
upper limit of 25\,mJy for a channel spacing of 1.05\,km\,s$^{-1}$.
\keywords{Galaxies: individual: IC~342, Maffei~2 -- Galaxies: starburst -- 
          ISM: molecules -- masers -- Radio lines: ISM}}

\maketitle

\titlerunning{Water maser emission in IC~342}
\authorrunning{Tarchi et al.}


\section{Introduction}

Isotropic luminosities of extragalactic 22\,GHz $\rm H_{2}O$ masers span 
a huge range (see e.g. Ho et al.\ \cite{ho87}; Koekemoer et al.\ 
\cite{koeke95}; Braatz et al.\ \cite{braatz96}), from $\rm 10^{-2} 
- 10$~$\rm L_{\odot}$ (the `kilomasers' consistent with the most intense 
masers seen in our Galaxy) to 10 -- 500~$\rm L_{\odot}$ (the `megamasers', 
found in LINERs and Seyfert 2s) and up to 6000 $\rm L_{\odot}$ (the 
`gigamaser' in TXS\,2226-184). Interferometric observations, when available, 
show that the most luminous $\rm H_{2}O$ sources arise from the nuclear 
region of their host galaxies, from circumnuclear tori, from entrained 
material near the surface of the nuclear jet, or by ambient gas amplifying 
the continuum emission of the jet. The weaker masers often mark locations 
of high mass star formation (for M~33, IC~10, and M~82, see e.g. Huchtmeier 
et al.\ \cite{hucht78}; Henkel et al.\ \cite{henkel86}; Baudry \& Brouillet 
\cite{baundry96}). The nature of the kilomasers in NGC~253 and M~51 (see Ho 
et al.\ \cite{ho87}; Nakai \& Kasuga \cite{nakai88}) remained an enigma 
for more than a decade. Recently, however, it has been shown that the maser 
in M~51 is of nuclear origin (Hagiwara et al.\ \cite{hagi01}), suggesting that 
there might be a family of as yet unexplored `weak' H$_2$O sources located near the nuclear engine of Seyfert 2 and LINER galaxies. 

Extragalactic H$_2$O masers are preferentially detected in nearby
galaxies that are bright in the infrared (Braatz et al.\ \cite{braatz97}).
While nuclear masers are of obvious interest, non-nuclear kilomasers 
are also important for a number of reasons: these sources allow us
to pinpoint sites of massive star formation, to measure the velocity
vectors of these regions through VLBI proper motion studies, and to 
determine true distances through complementary measurements of proper 
motion and radial velocity (e.g. Greenhill et al.\ \cite{greenhill93}). 
We have therefore observed the nearby spiral galaxies IC~342 and Maffei~2, both of which exhibit prominent nuclear bars and strong molecular, infrared, and radio 
continuum emission (e.g. Hurt \& Turner\ \cite{hurt91}; Hurt et al.\ 
\cite{hurt93}; Turner \& Ho\ \cite{turner94}; Henkel et al.\ \cite{henkel00}; 
Meier et al.\ \cite{meier00}; Meier \& Turner\ \cite{meier01}; Schulz et al.\ 
\cite{schulz01}). In the following we report the results of our observations. 
 
\section{The observation and image processing}

\paragraph{\bf {Effelsberg}} The $6_{16} - 5_{23}$ line of H$_2$O (rest 
frequency: 22.23508\,GHz) was observed with the 100-m telescope of the 
MPIfR at Effelsberg\footnote{The 100-m telescope at Effelsberg is operated by the Max-Planck-Institut f\"ur Radioastronomie (MPIfR) on behalf of the Max-Planck-Gesellschaft (MPG).} toward IC~342 and Maffei~2. The full width
to half power beamwidth was 40\arcsec. A dual channel HEMT receiver 
provided system temperatures of $\sim$200\,K on a main beam brightness 
temperature scale. The observations were carried out in a dual beam switching 
mode with a beam throw of 2\arcmin\ and a switching frequency of $\sim$1\,Hz.
 Flux calibration was obtained by measurements of W3(OH) (for fluxes, see 
Mauersberger et al.\ \cite{mauer88}). Gain variations as a function of
elevation were taken into account and the calibration error is expected
to be no more than $\pm$10\%. The pointing accuracy was always better 
than 10\arcsec.

\paragraph{\bf {VLA}}

On May 12, 2001, IC~342 was observed with the Very Large Array\footnote{The National Radio Astronomy Observatory is a facility of the National Science Foundation operated under cooperative agreement by Associated Universities, Inc.} (VLA) in its B configuration. We observed with a single IF using full polarimetric 
information. The source 0538+498 was used as flux calibrator. The nearby point source 0224+671 was used for phase and bandpass calibration. `Referenced pointing' was performed during the observations (for 
details, refer to http://www.aoc.nrao.edu/vla/html/refpt.shtml).

The data were Fourier-transformed using natural weighting to produce a $\rm 
2048 \times 2048 \times 103$ spectral line data cube, encompassing the 
central 100\arcsec\ of IC~342. The 103 channels used, out of the 128 observed, 
cover a range in velocity of $\sim$8\,km\,s$^{-1}$ centered at 16\,km\,s$^{-1}$ 
LSR (the velocity of the line detected at Effelsberg). No continuum subtraction 
was needed. The data were deconvolved using the CLEAN algorithm (H\"ogbom 
\cite{hoegbom74}). The restoring beam is $\rm 0\farcs4 \times 0\farcs3$ 
and the rms noise per channel is $\sim$10\,mJy, consistent with the expected 
thermal noise. 
 
\section{Results}

\subsection{Time variability}

Our single-dish observations towards Maffei~2 yielded no detection, with a 5$\sigma$ upper 
limit of 25\,mJy (channel spacing: 1.05\,km\,s$^{-1}$; velocity range: 
--250\,km\,s$^{-1}$ $<$ $V$ $<$ +230\,km\,s$^{-1}$; epoch: Apr 3, 2001; 
position: $\alpha_{2000}$ = 02$^{\rm h}$ 41$^{\rm m}$ 55$\fs$2, $\delta_{2000}$ 
= +59$\degr$ 36$\arcmin$ 11$\arcsec$).

\begin{figure}[h]
\centering
\resizebox{8.0cm}{!}{\includegraphics{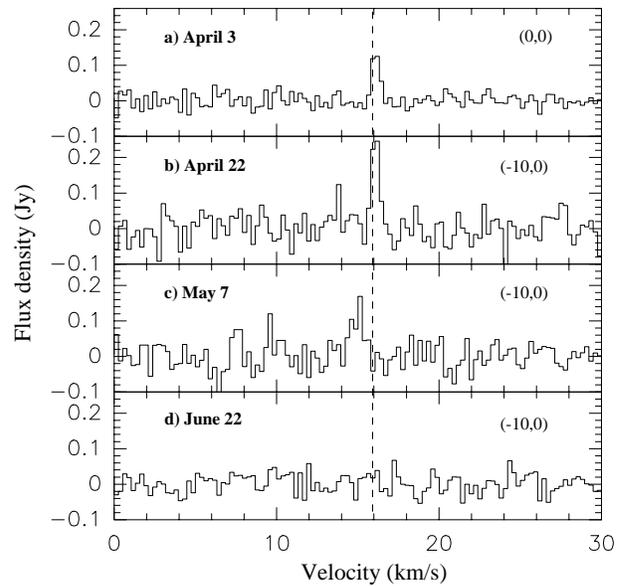}}
\caption{The H$_2$O maser feature observed with high velocity resolution 
(channel spacings after smoothing four contiguous channels are 0.26\,km\,s$^{-1}$) 
on {\bf {a)}} April 3, {\bf {b)}} April 22, {\bf {c)}} May 7, and {\bf {d)}} June 22. 
The first spectrum has been taken at the (0\arcsec,0\arcsec) position, the others 
at (--10\arcsec,0\arcsec) relative to the position given in footnote `a' of Table 1. 
The dashed line indicates $V_{\rm LSR}$ = 16\,km\,s$^{-1}$.}
\label{flare}
\end{figure}

\begin{figure*}[h]
\centering
\resizebox{16.0cm}{!}{\includegraphics{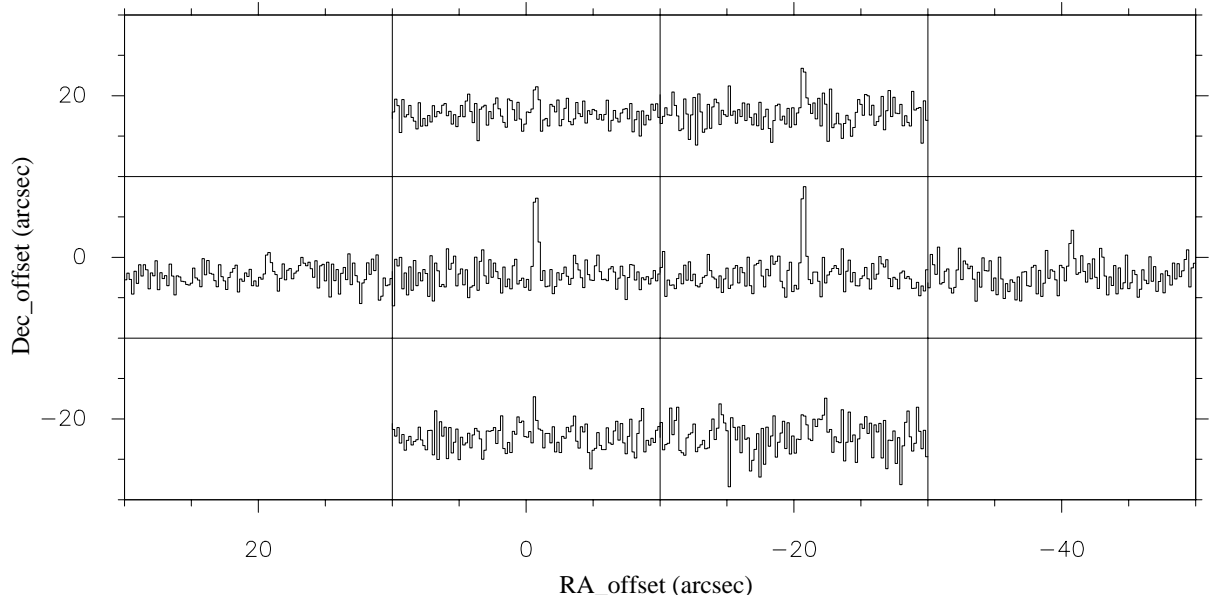}}
\caption{H$_2$O spectra obtained toward the central region of IC~342. 
Positions are offsets relative to \mbox{$\rm \alpha_{2000} = 03^{h} 46^{m} 
48\fs6$} and \mbox{$\delta_{2000} = +68\degr 05\arcmin 46\arcsec$}. Note 
that the spacing between two individual spectra (20\arcsec) is approximately 
half of the size of the 100-m Effelsberg telescope beam at 22\,GHz (40\arcsec). 
Averaging four contiguous channels, the spectra have been smoothed to a 
spacing of 0.26~km\,s$^{-1}$.}
\label{map}
\end{figure*}

On April 2, 2001, we obtained the first 
definite detection of water vapor emission in IC~342\footnote{Water masers in 
IC~342 were previously reported by Huchtmeier et al.\ (\cite{hucht78}) well 
outside the nuclear region. However, their spectrum from IC~342-4 was not shown 
and its velocity (--54\,km\,s$^{-1}$) is highly `forbidden' (there is no \ion{H}{i} at this velocity; see e.g. Crosthwaite 
et al.\ \cite{crost00}), while the spectral profile from IC~342-3, peaking at 
--52\,km\,s$^{-1}$, looks tentative. If the feature is real, an association 
with the galactic Perseus arm cannot be excluded.} (see also Sect.\,4.1). During 
the next night the detection was confirmed with a velocity resolution sufficient 
to resolve the line profile (Fig.\,\ref{flare}a). The detected feature lies at 
$V_{\rm LSR}$ = 16\,km\,s$^{-1}$ and has a linewidth of $\sim$0.5\,km\,s$^{-1}$; 
on April 2, no other component was seen at velocities --175\,km\,s$^{-1}$ $<$ $V$ 
$<$ +310\,km\,s$^{-1}$ (channel spacing: 1.05\,km\,s$^{-1}$; 5$\sigma$ noise 
level: 16\,mJy). A high line intensity ($\sim$100\,mJy) and good weather 
conditions allowed us to map the emitting region. Fig.\,\ref{map} shows the 
spectra taken towards six positions near the center of the galaxy. On April 
22, observations were performed towards the offset position (--10\arcsec,0\arcsec), 
close to the peak position inferred by the map. The flux density had almost 
doubled (compare Figs.\,\ref{flare}a and b) which is more than the expected 20\% 
increase for a point source located $\sim$10\arcsec\ off our previously observed 
(0\arcsec,0\arcsec) position. Further measurements were performed on May 7 
(Fig.\,\ref{flare}c). The peak and integrated intensities are smaller and the velocity 
of the line is blue-shifted by $\sim$1\,km\,s$^{-1}$. Unfortunately, no emission 
above $\sim$30\,mJy (3$\sigma$ level; 0.08\,km\,s$^{-1}$ channel spacing) was 
detected in the 22~GHz VLA B-array data taken on May 12. This fading by at
least a factor of 3 within 5 days implies a size scale of $\la$\,$1.5\times10^{16}$\,cm
($\la$\,900\,AU) or $\la$\,0.5\,mas at a distance of 1.8\,Mpc (McCall et al.\
\cite{mccall89}; see also Sect.\,4.3); the corresponding brightness temperature 
is $\ga$10$^{9}$\,K. The most recent spectrum was obtained on June 22 at Effelsberg (Fig.\,\ref{flare}d). 
Confirming the VLA result, no maser signal was seen above 30\,mJy (3$\sigma$; 
channel spacing: 1.05\,km\,s$^{-1}$).

Offset positions, on-source integration times, Gaussian fit parameters, and 
H$_2$O luminosities assuming isotropic emission (see e.g. Henkel et al.\ 
\cite{henkel98}) are given in Table~\ref{fluxes}. 

\begin{table*}
\label{fluxes}
\caption{H$_2$O in IC~342: observations details and line parameters}
\scriptsize
\begin{center}
\begin{tabular}{lccccccccc}
\hline
\\
Obs. Date & $(\Delta \alpha, \Delta \delta)^{\rm a)}$ & Tel. & 
$t_{\rm obs}^{\rm b)}$  & BW & Vel. Res.& $S_{\rm peak}^{\rm c)}$ & $V_{\rm LSR}^{\rm d)}$ & 
$\Delta V_{1/2}$ & Luminosity$^{\rm e)}$ \\
\multicolumn{1}{c}{} & 
\multicolumn{1}{c}{(\arcsec,\arcsec)} & 
\multicolumn{1}{c}{} & 
\multicolumn{1}{c}{(min)} &
\multicolumn{1}{c}{(MHz)} &
\multicolumn{1}{c}{(km\,s$^{-1}$)} & 
\multicolumn{1}{c}{(Jy)} & 
\multicolumn{2}{c}{(km\,s$^{-1}$)} & 
\multicolumn{1}{c}{(L$_{\odot}$)} \\
\\
\hline
\\
02 Apr 2001$^{\rm f)}$ & (0,0)        & EFF &180 & 40 & 1.05 &  --           & 16.16 (0.03) &  --         & (8.3$\pm$0.6)$\times$10$^{-3}$\\
03 Apr 2001            & (--40,0)     & EFF & 30 & 20 & 0.07 & 0.07$\pm$0.03 & 16.15 (0.09) & 0.61 (0.28) & (3.2$\pm$1.0)$\times$10$^{-3}$\\
03 Apr 2001            & (--20,--20)  & EFF & 20 & '' & '' & $<$0.07       &              &             &  $<$3$\times$10$^{-3}$        \\
03 Apr 2001            & (--20,0)     & EFF & 30 & '' & '' & 0.18$\pm$0.03 & 16.08 (0.03) & 0.49 (0.06) & (6.5$\pm$0.7)$\times$10$^{-3}$\\
03 Apr 2001            & (--20,20)    & EFF & 30 & '' & '' & 0.10$\pm$0.03 & 16.07 (0.06) & 0.53 (0.13) & (3.9$\pm$0.8)$\times$10$^{-3}$\\
03 Apr 2001            & (0,--20)     & EFF & 30 & '' & '' & 0.06$\pm$0.02 & 16.06 (0.08) & 0.46 (0.14) & (2.0$\pm$0.6)$\times$10$^{-3}$\\
03 Apr 2001            & (0,0)        & EFF & 30 & '' & '' & 0.17$\pm$0.02 & 16.14 (0.02) & 0.58 (0.05) & (7.3$\pm$0.6)$\times$10$^{-3}$\\
03 Apr 2001            & (0,20)       & EFF & 30 & '' & '' & 0.06$\pm$0.04 & 16.15 (0.06) & 0.40 (0.20) & (1.8$\pm$0.6)$\times$10$^{-3}$\\
03 Apr 2001            & (20,0)       & EFF & 30 & '' & '' & 0.04$\pm$0.02 & 16.14 (0.13) & 0.64 (0.22) & (1.9$\pm$0.7)$\times$10$^{-3}$\\
22 Apr 2001            & (--10,0)     & EFF &  8 & '' & '' & 0.30$\pm$0.06 & 16.11 (0.04) & 0.54 (0.08) &(12.0$\pm$1.5)$\times$10$^{-3}$\\
07 May 2001            & (--10,0)     & EFF & 15 & '' & '' & 0.17$\pm$0.05 & 14.98 (0.08) & 0.81 (0.18) &(10.3$\pm$1.9)$\times$10$^{-3}$\\
12 May 2001            &              & VLA & 120 & 0.781 & 0.08  & $<$0.03        &              &             &  $<$1.2$\times$10$^{-3}$      \\
22 Jun 2001            &(--10,0)+(0,0)& EFF & 35 & 40 & 1.05 & $<$0.03        &              &             &  $<$1.7$\times$10$^{-3}$      \\
\\
\hline
\end{tabular}
\end{center}
a) The position offset is relative to $\alpha_{2000}$ = 03$^{\rm h}$ 46$^{\rm m}$ 48\fs6 
and $\delta_{2000}$ = +68\degr 05\arcmin 46\arcsec \\
b) $t_{\rm obs}$ (Effelsberg 22\,GHz H$_2$O integration time) includes time
for on- and off-source integration \\
c) Peak fluxes calculated from integrated line intensities divided by linewidths (both
obtained from Gaussian fits using the data reduction software package `CLASS'); if no 
signal is detected, 3$\sigma$ limits are given for a channel spacing of 0.53\,km\,s$^{-1}$ 
(Apr.~3), 0.08\,km\,s$^{-1}$ (May~12), and 1.05\,km\,s$^{-1}$ (June~22). Since the data 
taken on April 3 were used for the computation of the position of the maser, where only 
(presumably negligible) relative calibration errors are relevant, stated errors do not 
include absolute calibration errors ($\pm$10\%; see Sect.\,2). \\
d) LSR = Local Standard of Rest \\
e) Obtained via [$L_{\rm H_2O}$/L$_{\odot}$] = 0.023 $\times$ 
[$\int{S\,{\rm d}V}$/Jy\,km\,s$^{-1}$] $\times$ [$D$/Mpc]$^{2}$, $D$ = 1.8\,Mpc.
Corresponding luminosity limits refer to $\Delta V_{1/2}$ = 0.55\,km\,s$^{-1}$,
the approximate width of the H$_2$O feature. Errors do not include the estimated 
calibration uncertainty of $\pm$10\% (see Sect.\,2 and footnote c). \\
f) Velocity resolution too coarse to determine meaningful values for the 
peak flux density and linewidth \\
\end{table*}

\subsection{Maser position}

Fitting a synthetic Gaussian to the data taken on April 3 (see Table 1 and
Fig.\,\ref{map}), i.e. minimizing the sum of the difference squared between 
calculated and observed peak and integrated flux densities, we can obtain 
an accurate position of the emitting region: $\alpha_{2000}$ = 03$^{\rm h}$ 
46$^{\rm m}$ 46\fs3, $\delta_{2000}$ = +68\degr 05\arcmin 46\arcsec. This 
is 13\arcsec\ to the west of our center position (see Table 1) that coincides
with the optical nucleus and the 2$\mu$m peak (van der Kruit\ \cite{kruit73}; 
Becklin et al.\ \cite{becklin80}). The accuracy of the derived maser position 
is affected by the following uncertainties: 

{\it Relative pointing accuracy:} On April 3, the average pointing correction 
was $\sigma_{\rm rpoi}$$\sim$2\arcsec. Using the nearby 
ultra-compact \ion{H}{ii} region W3(OH) as pointing source slightly degraded 
this pointing accuracy because of the presence of a strong H$_2$O maser 6\arcsec\ to the west 
(W3(H$_2$O); e.g. Reid et al.\ \cite{reid95}). The error introduced by the non-negligible flux contribution of the 
maser is $\sigma_{\rm W3(OH)}$ = 1\farcs2.

{\it Absolute pointing accuracy:} When moving the Effelsberg antenna from W3(OH)
to IC~342, there may be an additional pointing offset caused by the change in
azimuth and elevation. On April 3, relatively large telescope movements led to a pointing shift of 2\arcsec\ in azimuth and 4\arcsec\ in elevation. Considering the improvement in weather conditions when subsequently mapping IC~342, and accounting for the much smaller distance between target (IC~342) and pointing source (W3(OH)), we estimate the absolute pointing error to be $\sigma_{\rm apoi}$ = 2\farcs5.  

{\it Spectral noise:} For data taken during a single night, relative calibration
should be excellent. The relevant factor affecting line intensity ratios is 
thus noise. Assuming a Gaussian beamshape (beamwidth: 40\arcsec), resulting peak 
and integrated H$_2$O flux densities were calculated for various maser positions. 
Agreement with the observed intensity limit at the (--20\arcsec,--20\arcsec) position was taken 
as an additional check. The resulting positional error is $\sigma_{\rm map}$ = 4\arcsec. 
 
The assumption that all errors can be statistically added yields a total 
positional uncertainty of \mbox{$\sigma_{\rm tot} = \sqrt{\sigma_{\rm rpoi}^{2}
+ \sigma_{\rm apoi}^{2} + \sigma_{\rm W3(OH)}^{2} + \sigma_{\rm map}^{2}} \sim\ 
5\arcsec$}. 

\section{Discussion}

\subsection{Galactic versus extragalactic H$_2$O emission}

\begin{figure}[h]
\centering
\resizebox{8.0cm}{!}{\includegraphics{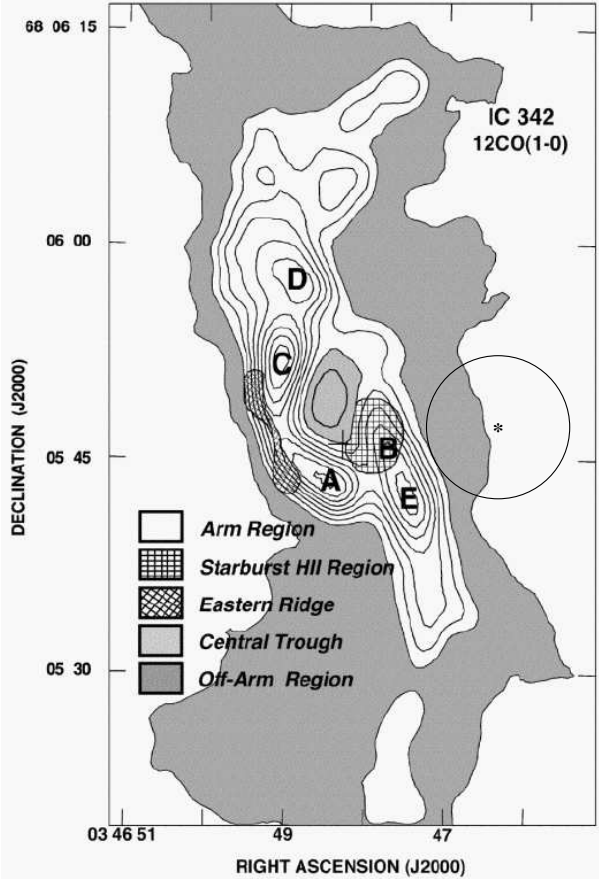}}
\caption{A map of integrated $^{12}$CO $J$ = 1--0 emission from the central region 
of IC~342 (Meier et al.\ \cite{meier00}). The asterisk indicates the location of
the water maser with the circle marking the positional uncertainty (Sect.\,3.2). 
The letters label the five giant molecular clouds identified by Downes et al.\ 
(\cite{downes92}).}
\label{mappaco}
\end{figure}

IC~342 is located at a galactic longitude of 138\degr\ and latitude 10\degr,
i.e. behind the northern outskirts of the Perseus arm, the Cam OB1 association,
and a Local Arm of the Milky Way (e.g. Digel et al.\ \cite{digel96}; Dame et 
al.\ \cite{dame01}). Local Standard of Rest (LSR) velocities of the Perseus 
arm range from --60 to --30\,km\,s$^{-1}$, consistent with the maser features 
reported by Huchtmeier et al.\ (\cite{hucht78}), but significantly removed from the radial velocity of the maser reported here. For the Cam OB1 association and the Local Arm, 
Digel et al.\ (\cite{digel96}) find velocities up to --5 and +7\,km\,s$^{-1}$, 
still below the measured maser velocity. Furthermore, only a few of the hundreds of 
known galactic H$_2$O masers have been detected at $b^{\rm II}$$\sim$10\degr\ 
(e.g. Valdettaro et al.\ \cite{valdettaro01}). All of these arguments support an extragalactic origin of the maser. In addition, the IC~342 isovelocity contour at 
$V_{\rm LSR}$ $\sim$ 16\,km\,s$^{-1}$, obtained from CO interferometric data, 
is consistent with a maser location 10 -- 15\arcsec\ west of its nucleus 
(e.g. Lo et al.\ \cite{lo84}; Sakamoto et al.\ \cite{sakamoto99}). This 
agrees with the position determined in Sect.\,3.2. We therefore conclude that 
the H$_2$O source is associated with the central region of IC~342 (for an 
illustration, see Fig.~\ref{mappaco}).

\subsection{Search for an optical counterpart}

Arising from the central region of IC~342, but being displaced from the nucleus, 
the H$_2$O maser is likely associated with a prominent star forming region.

\begin{figure}[h]
\centering
\resizebox{8.0cm}{!}{\includegraphics{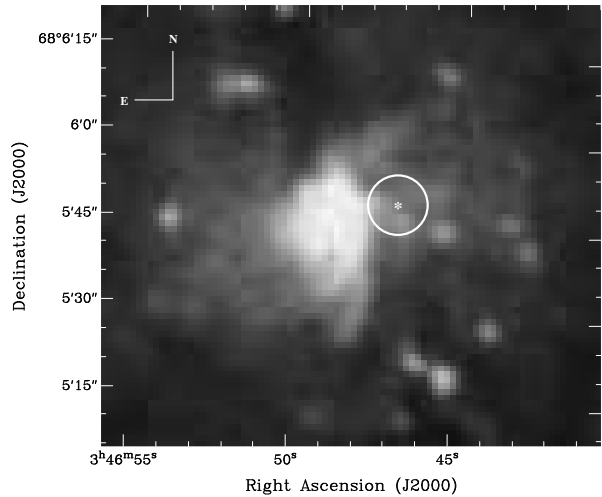}}
\caption{XDSS B-band image of the central region of IC~342. The asterisk indicates 
the water maser emitting position. The radius of the circle indicates its
positional error as outlined in Sect.\,3.2.}
\label{mappadss}
\end{figure}

Fig.~\ref{mappadss} shows an optical B-band image of the central region of IC~342, taken from the XDSS\footnote{The Digitized Sky Surveys were produced at the Space Telescope Institute under U.S. Government grant NAGW-2166. The images of these surveys are based on photographic data obtained using the Oschin Schmidt Telescope on Palomar Mountain and the UK Schmidt Telescope. The plates were processed into the present compressed digital form with the permission of these institutions.}. 
The maser emitting region (white circle) coincides with an arc-like structure 
extending E-S and is associated with a chain of sources that appear to be 
\ion{H}{ii} regions. The presence of optically bright regions, likely sites of star formation activity, close to the maser position is confirmed by an $\rm H\alpha$ ($\lambda \rm 6563\:\AA$) HST WFPC2 picture\footnote{Some of the data mentioned in this paper were obtained from the Multimission Archive at the Space Telescope Institute (MAST). STScI is operated by the Association of Universities for Research in Astronomy, Inc. under NASA contract NAS 5-26555. Support for MAST for non-HST data is provided by the NASA Office of Space Science via grant NAG5-7584 and by other grants and contracts.} (see also Meier et al.\ \cite{meier00}).

\subsection{A comparison with other maser sources}

Although seen almost face-on (inclination $i$$\sim$25\degr; e.g. Sage \& 
Solomon \cite{sage91}), IC~342 is believed to resemble the Milky Way in many 
ways. At the commonly assumed $D$ $\sim$ 2\,Mpc (but see Krismer et al.\ 
\cite{krismer95} for a larger distance), the giant molecular clouds near 
the center of IC~342 have similar linear sizes to the Sgr~A and Sgr~B2 
clouds in our Galaxy. The inner 400~pc of IC~342 and the Galaxy have 
almost the same far infrared luminosity and the 2$\mu$m luminosities 
indicate similar stellar masses (Downes et al.\ \cite{downes92}). 

Does this similarity extend to the 22\,GHz maser emission? While there are
a number of H$_2$O maser sources in the nuclear region of the Milky Way
(G{\"u}sten \& Downes\ \cite{guesten83}; Taylor et al.\ \cite{taylor93};
Yusef-Zadeh \& Mehringer\ \cite{yusef95}), the emission is dominated by 
Sgr\,B2 (e.g. Waak \& Mayer \cite{waak74}; Morris \cite{morris76}; Genzel 
et al.\ \cite{genzel76}; Elmegreen et al.\ \cite{elmegreen80}; Kobayashi 
et al.\ \cite{koba89}). Like the H$_2$O maser in IC~342, Sgr\,B2 is displaced 
from the galactic nucleus. The integrated luminosity of its numerous spatial 
components is $L_{\rm H_2O}$ $\sim 5\times10^{-3}$ L$_{\odot}$, again not
inconsistent with the data from IC~342. The lineshapes are, however, quite different: 
While Sgr\,B2 is characterized by a variety of narrow components that have 
flux densities of a few hundred Jy, the velocity component we have detected in IC~342 is stronger than any related feature by at least an order of 
magnitude. 

Occasionally, prominent galactic star forming regions show narrow 
($\sim$0.5\,km\,s$^{-1}$) flaring components that are exceptionally bright. 
Such flares were observed in W\,31\,A and W\,49 (Liljestr{\"o}m et al.\ 
\cite{lilje89}; Lekht et al.\ \cite{lekht95}). The prototype of these 
flares is the 8\,km\,s$^{-1}$ super maser in Orion-KL (e.g. Garay et al.\ 
\cite{garay89}). During several months, the narrow highly linearly polarized 
maser reached flux densities in excess of 5\,MJy. Surpassing the flux of 
any other velocity component by more than an order of magnitude and 
reaching a peak luminosity of $L_{\rm H_2O}$ $\sim$ 10$^{-2}$\,L$_{\odot}$, 
the feature seems to be similar to that seen in IC~342. There is another even 
more impressive narrow flaring maser, which dwarfed neighboring
features: the \mbox{--324\,km\,s$^{-1}$} component in the nearby irregular galaxy 
IC\,10 (Becker et al.\ \cite{becker93}; Argon et al.\ \cite{argon94}; Baan \& Haschick\ \cite{baan94}). 
This prominent flare, reaching a luminosity of $\sim$1\,L$_{\odot}$ 
($\sim$100\,Jy; $D$ $\sim$ 0.8\,Mpc), lasted for a few years. As unpublished 
spectra from Effelsberg show, the lineshape of the maser also resembles that 
of the flaring feature in Orion-KL. 

While it is not farfetched to assume that the flares in IC~342, IC~10, and Orion-KL are caused by the same physical phenomenon (but see Argon et al.\ \cite{argon94}), 
the actual situation remains unclear. For Orion-KL, observational constraints 
are quite strict. There is no significant background continuum source and the 
high intensity seems to suggest saturated emission and a collisional pump 
(e.g. Garay et al.\ \cite{garay89}). The small linewidth, however, appears 
to be incompatible with such a scenario (Nedoluha \& Watson\ \cite{nedoluha91}). 
Solving this apparent inconsistency requires either very small beaming angles, 
which are readily obtained by a chance alignment of two masing clouds (e.g. 
Deguchi \& Watson\ \cite{deguchi89}; Elitzur et al.\ \cite{elitzur91}), or 
the inclusion of efficient photon scattering, which can account for linewidths 
$\sim$0.5\,km\,s$^{-1}$ even in the case of saturation (e.g. Elitzur\ 
\cite{elitzur90}). 

While luminosity, linewidth, and flux variations are reminiscent of the 
8\,km\,s$^{-1}$ super maser in Orion-KL, our H$_2$O profiles show a
significant velocity shift between April 22 and May 7 (Figs.~\ref{flare}b and c)
that has not been seen in the Orion-KL flaring component. The shift, $\sim$1\,km\,s$^{-1}$
(Table 1; see also Sect.\,3.1), is too large to be explained by a change 
in the relative intensities of the three main hyperfine components (see e.g. 
Fiebig \& G{\"u}sten\ \cite{fiebig89}). A change with time in the peak velocity could originate from a variation in the relative intensity of different emitting regions leading to the swapping-over of velocity peaks (see e.g.\ Wu et al.\ \cite{wu99}). Such intensity fluctuations have been found to be sometimes ``correlated'' with observed time scales ranging from some days up to weeks (intrinsic time scales are more difficult to deduce because they strongly depend on the degree of saturation of the masers; see e.g. Genzel \& Downes\ \cite{genzel77}; Rowland \& Cohen\ \cite{rowland86}). Time scales equal or greater than some months are also predicted for maser flux density variations (distinct from flares) due to refractive interstellar scintillations (Simonetti et al.\ \cite{simonetti93}). However, a simpler model for explaining both the flare and the shift is, in our opinion, that the latter has a kinematic origin. 
Adopting the previously outlined scenario of 
a chance alignment of two masing clouds along the line-of-sight, motion 
of the foreground relative to the background cloud along the plane of the 
sky and velocity structure in the foreground cloud could explain the observed 
data. A velocity gradient in the foreground cloud would then first shift 
the line velocity; once velocities are reached that are not matched within the background cloud, the flux density of the maser rapidly drops. Assuming 
that the distance between these clouds is $\ll$1.8\,Mpc and that their relative 
velocity in the plane of the sky is at the order of 100\,km\,s$^{-1}$, this 
implies a cloud velocity gradient of up to 1\,km\,s$^{-1}$/AU during the time
the source was monitored.

An almost identical scenario was proposed by Boboltz et al.\ (\cite{boboltz98}) to account for the velocity shift of the flaring water maser component at --66\,km\,s$^{-1}$ in W49N. The similarity in velocity shift ($\sim$ 0.5\,km\,s$^{-1}$) and time scale (58 days) of the event in W49N w.r.t the one in IC~342, indicates a common origin.  

\subsection{Extragalactic maser detection rates}

While typical searches for extragalactic maser sources have only yielded detection
rates between zero (e.g. Henkel et al.\ \cite{henkel98}) and a few percent
(e.g. Henkel et al.\ \cite{henkel84}; Braatz et al.\ \cite{braatz96}), there
exists one sample with detection rates $>$10\%: These are the northern 
($\delta$ $>$ --30\degr) extragalactic IRAS (Infrared Astronomy Satellite) 
point sources with 100$\mu$m fluxes in excess of 50\,Jy (for a source list,
see Henkel et al.\ \cite{henkel86}; Maffei 2 with $S_{\rm 100\mu m}$ $\sim$ 200\,Jy
should be added to the list). There is a total of 44 galaxies, two ultraluminous 
galaxies at intermediate distances (NGC\,3690 and Arp\,220) and 42 nearby sources 
($V$ $<$ 3000\,km\,s$^{-1}$). Among these, seven (16\%) are known to contain 
H$_2$O masers in their nuclear region. Two of these contain megamasers 
(NGC~1068 and NGC~3079), two are possibly nuclear kilomasers (NGC~253 and M~51; 
for details, see Sect.\,1), and three are associated with prominent sites 
of star formation (IC~10, IC~342, M~82). Among the subsample of 19 sources 
with 100$\mu$m fluxes in excess of 100\,Jy, five were so far detected in 
H$_2$O, yielding a detection rate in excess of 20\%. Since few deep integrations 
have been obtained toward these sources, more H$_2$O detections can be expected from 
this promising sample in the near future.

\section{Conclusions}

Having detected water vapor emission in the galaxy IC~342, we have obtained
the following main results:

\begin{enumerate}

\item The maser arises from a location 10--15\arcsec\ to the west of the center
of the galaxy. It is thus a maser associated with a powerful star forming region
at a projected distance of $\sim$100\,pc ($D$ = 1.8\,Mpc) from the nucleus.

\item Luminosity (10$^{-2}$\,L$_{\odot}$), linewidth (0.5\,km\,s$^{-1}$), and rapid 
flux density variations are reminiscent of the flaring 8\,km\,s$^{-1}$ super maser 
feature in Orion-KL. It seems that we have observed an exceptional outburst of a maser that is usually well
below the detection threshhold.

\item Time variability, if intrinsic, yields a maser size of $\la 1.5\times10^{16}$\,cm$^{-3}$
($\la$0.5\,mas), and a brightness temperature of $\ga$10$^{9}$\,K. 

\item A velocity shift of 1\,km\,s$^{-1}$ within 16 days and a subsequent decrease 
in flux density by at least a factor of three within 5 days can be explained by a 
chance alignment of two dense molecular clouds along the line-of-sight, both of 
them associated with IC~342.

\item Among galaxies with IRAS point source flux densities of $S_{\rm 100\mu m}$ $>$
50\,Jy, 16\% are now known to contain masers in their inner regions. Since deep 
H$_2$O integrations have been obtained towards only a few of them, more detections from
this sample can be expected in the near future.

\end{enumerate}

\begin{acknowledgements}

We wish to thank P. Diamond for critically reading the manuscript and Nikolaus Neininger for useful discussion. We are also endebted to the operators at the 100-m telescope, and to Michael Rupen and the NRAO analysts, for their cheerful assistance with the observations.

\end{acknowledgements}


\begin{thebibliography}{}

\bibitem[1994]{argon94}
Argon A.L., Greenhill L.J., Moran J.M., Reid M.J., et al.\, 1994, ApJ 422, 586

\bibitem[1994]{baan94}
Baan W.A., Haschick A.D., 1994, ApJ 424, L33

\bibitem[1996]{baundry96}
Baudry A., Brouillet N., 1996, A\&A 316, 188

\bibitem[1993]{becker93}
Becker R., Henkel C., Wilson T.L., Wouterloot J.G.A., 1993, A\&A 268, 483 

\bibitem[1980]{becklin80}
Becklin E.E., Gatley I., Matthews K., Neugebauer G., et al.\, 1980, ApJ 236, 441

\bibitem[1998]{boboltz98}
Boboltz D.A., Simonetti J.H., Dennison B., Diamond P.J., et al.\, 1998, ApJ 509, 256

\bibitem[1996]{braatz96}
Braatz J.A., Wilson A.S., Henkel C., 1996, ApJS 106, 51

\bibitem[1997]{braatz97}
Braatz J.A., Wilson A.S., Henkel C., 1997, ApJS 110, 321

\bibitem[2000]{crost00}
Crosthwaite L.P., Turner J.L., Ho P.T.P., 2000, AJ 119, 1720

\bibitem[2001]{dame01}
Dame T.M., Hartmann D., Thaddeus P., 2001, ApJ 547, 792

\bibitem[1989]{deguchi89}
Deguchi S., Watson W.D., 1989, ApJ 340, L17

\bibitem[1996]{digel96}
Digel, S.W., Lyder, D.A., Philbrick, A.J., Puche, D., Thaddeus, P., 1996, ApJ 458, 561

\bibitem[1992]{downes92}
Downes D., Radford S.J.E., Guilloteau S., Gu{\'e}lin M., et al., 1992, A\&A 262, 424

\bibitem[1990]{elitzur90}
Elitzur M., 1990, ApJ 350, L7

\bibitem[1991]{elitzur91}
Elitzur M., McKee C.F., Hollenbach D.J., 1991, ApJ 367, 333 

\bibitem[1980]{elmegreen80}
Elmegreen B.G., Genzel R., Moran J.M., Reid M.J., Walker R.C., 1980, ApJ 241, 1007

\bibitem[1989]{fiebig89}
Fiebig D., G{\"u}sten R., 1989, A\&A 214, 333

\bibitem[1989]{garay89}
Garay G., Moran J.M., Haschick A.D., 1989, Apj 338, 244

\bibitem[1977]{genzel77}
Genzel R., Downes D., 1977, A\&AS 30, 145

\bibitem[1976]{genzel76}
Genzel R., Downes D., Bieging J., 1976, MNRAS 177, 101

\bibitem[1993]{greenhill93}
Greenhill L.J., Moran J.M., Reid M.J., Menten K.M., Hirabatashi H., 1993, ApJ 406, 482 

\bibitem[1983]{guesten83}
G{\"u}sten R., Downes D., 1983, A\&A 117, 343

\bibitem[2001]{hagi01}   
Hagiwara Y., Henkel C., Menten K., Nakai N., 2001, ApJ 560, L37 

\bibitem[1984]{henkel84}
Henkel C., G{\"u}sten R., Downes D., Thum C., Wilson T.L., Biermann P., 1984, A\&A 141, L1

\bibitem[1986]{henkel86}
Henkel C., Wouterloot J.G.A., Bally J., 1986, A\&A 155, 193

\bibitem[1998]{henkel98}
Henkel C., Wang Y.P., Falcke H., Wilson A.S., Braatz J.A., 1998, A\&A 335, 463

\bibitem[2000]{henkel00}
Henkel C., Mauersberger R., Peck A.B., Falcke H., Hagiwara Y., 2000, A\&A 361, L45

\bibitem[1987]{ho87}
Ho P.T.P., Martin R.N., Henkel C., 1987, ApJ 320, 663

\bibitem[1974]{hoegbom74}
H\"ogbom J.A., 1974, A\&AS 15, 417

\bibitem[1978]{hucht78}
Huchtmeier W., Witzel A., K{\"u}hr H., et al., 1978, A\&A 64, L21

\bibitem[1991]{hurt91}
Hurt R.L., Turner, J.L., 1991, ApJ 377, 434

\bibitem[1993]{hurt93}
Hurt R.L., Turner J.L., Ho P.T.P., Martin R.N., 1993, ApJ 404, 602

\bibitem[1989]{koba89}
Kobayashi H., Ishiguro M., Chikada Y., Ukita N., et al.\, 1989, PASJ 41, 141

\bibitem[1995]{koeke95}
Koekemoer A.M., Henkel C., Greenhill L.J., et al., 1995, Nat 378, 697

\bibitem[1995]{krismer95}
Krismer M., Tully R.B., Gioa M., 1995, AJ 110, 1584

\bibitem[1995]{lekht95}
Lekht E.E., Mendoza-Torres E., Sorochenko R.L., 1995, ApJ 443, 222

\bibitem[1989]{lilje89}
Liljestr{\"o}m T., Mattila K., Toriseva M., Anttila R., 1989, A\&AS 79, 19

\bibitem[1984]{lo84}
Lo K.-Y., Berge G.L., Claussen M.J., et al., 1984, ApJ 282, L59

\bibitem[1988]{mauer88}
Mauersberger R., Wilson T.L., Henkel C., 1988, A\&A 201, 123

\bibitem[1989]{mccall89}
McCall M.L., 1989, AJ 97, 1341

\bibitem[2000]{meier00}
Meier D.S., Turner J.L., Hurt R.L., 2000, ApJ 531, 200

\bibitem[2001]{meier01}
Meier D.S., Turner J.L., 2001, ApJ 551, 687

\bibitem[1976]{morris76}
Morris M., 1976, ApJ 210, 100

\bibitem[1988]{nakai88}
Nakai N., Kasuga T., 1988, PASJ 40, 139

\bibitem[1991]{nedoluha91}
Nedoluha G.E., Watson W.D., 1991, ApJ 367, L63

\bibitem[1995]{reid95}
Reid M.J., Argon A.L., Masson C.R., Menten K.M. et al., 1995, ApJ 443, 238

\bibitem[1986]{rowland86}
Rowland P.R., Cohen R.J., 1986, MNRAS 220, 233

\bibitem[1991]{sage91}
Sage L.J., Solomon P.M., 1991, ApJ 380, 392

\bibitem[1999]{sakamoto99}
Sakamoto K., Okumura S.K., Ishizuki S., et al., 1999, ApJS 124, 403

\bibitem[2001]{schulz01}
Schulz A., G{\"u}sten R., K{\"o}ster B., Krause D., 2001, A\&A 371, 25

\bibitem[1993]{simonetti93}
Simonetti J.H., Diamond P.J., Uphoff J.A., Boboltz D., Dennison B., 1993, in {\it Astrophysical Masers}, Lecture Notes in Physics 412, eds. Clegg A.W. \& Nedoluha G.E., Springer Verlag, Berlin, p. 311

\bibitem[1993]{taylor93}
Taylor G.B., Morris M., Schulman E., 1993, AJ 106, 1978

\bibitem[1994]{turner94}
Turner J.L., Ho P.T.P., 1994, ApJ 421, 122

\bibitem[2001]{valdettaro01}
Valdettaro R., Palla F., Brand J., Cesaroni R., et al., 2001, A\&A 368, 845

\bibitem[1973]{kruit73}
van der Kruit P.C., 1973, A\&A 29, 249

\bibitem[1974]{waak74}
Waak J.A., Mayer J.A., 1974, ApJ 189, 67

\bibitem[1999]{wu99}
Wu Y.-F., Shi J.-R., Wang J.-Z., Jiang D.-R., Huang X.-Y., 1999, ChA\&A 23, 454

\bibitem[1995]{yusef95}
Yusef-Zadeh F., Mehringer D.M., 1995, ApJ 452, L17


\end{thebibliography}
\end{document}